\documentclass[10pt,twocolumn]{article}

\usepackage{ol2}    

 \usepackage{amssymb}    
 \usepackage{geometry}
\usepackage{graphicx}
\usepackage{epstopdf}
\usepackage[caption=false]{subfig}
 \usepackage{float}

\def\AOM {acousto-optic modulator}

\def\DVM {digital volt meter}
\def\DRO {dielectric resonator-oscillator}
\def\ea{\textit{et al.}}

\def\ICL {intercombination line}
\def\ICR {inverted crossover resonance}

\def\MOT {magneto-optical trap}

\def\PMT {photo-multiplier tube}

\def\wrt {with respect to}

\def\ARC {Australian Research Council} 

\newcommand{\degC}{$^{\circ}$C}

\newcommand{\uK}{$\mu$K}  
\newcommand{\us}{$\mu$s}

\newcommand{\fastT}{$^{1}S_{0}-\,^{1}P_{1}$}    
\newcommand{\coolingT}{$^{1}S_{0}-\,^{3}P_{1}$}  

\newcommand{\clockT}{$^{1}S_{0}-\,^{3}P_{0}$}

\newcommand{\clockTdash}{$^{1}S_{0} -\,^{3}P_{0}$} 

\newcommand{\si}{$\sim$}

\newcommand{\Yb}{$^{171}$Yb}

\begin{document}

  \twocolumn[ 

\title{Intercombination line frequencies in  $^{171}$Yb  validated with the clock transition}

\author{ Daniel M. Jones, Frank van Kann and John~J.~McFerran  }   %

\address{Department of Physics, The University of Western Australia, 6009 Crawley, Australia}  
\address{Corresponding author: john.mcferran@uwa.edu.au}  

\date{\today}

\begin{abstract}  
We have carried absolute frequency measurements  of the $(6s^{2})\,^{1}S_{0}$ $-$ $(6s6p)\,^{3}P_{1}$ transition in $^{171}$Yb (the intercombination line), where the spin-1/2 isotope yields two hyperfine lines. The measurements rely on sub-Doppler spectroscopy  to yield a discriminator to which a 556\,nm laser is locked.  
The frequency reference for the  optical frequency measurements is a high-quality quartz  oscillator steered to  the GNSS timescale 
that is  bridged with a frequency comb.  The  reference is  validated to $\sim3\times10^{-12}$ by spectroscopy on the  $^{1}S_{0}-\,^{3}P_{0}$ (clock) line in laser cooled and trapped $^{171}$Yb atoms.  From  the hyperfine separation between  the $F=1/2$ and $F=3/2$ levels of  $^{3}P_{1}$ we determine the hyperfine constant to be $A(^3P_1)= 3\,957\,833\,(28)$\,kHz.
   \end{abstract}   

    ] 


\section{Introduction}

The ytterbium \clockT\ line has  proven  to be an excellent choice for a frequency reference and time standard~\cite{Beloy2021, Piz2017, Nemitz2016, Bel2014}.    While not a contender for state-of-the-art optical clocks, the knowledge of the absolute frequencies of the \ICL s (ICLs) is  important for comparison with atomic structure calculations~\cite{Sch2021a,Dzu2010,Por1999,Liu1998a}.  
 Complex modelling of the Yb atom becomes robust when the calculated energy level separations agree  with measured values.   Outputs of the atomic structure calculations, such as hyperfine structure constants,  and isotope  shift parameters  become more reliable 
 when there is such agreement.  Also between calculated and measured hyperfine constants themselves.  Isotope  shift parameters have become particularly relevant in the recent work related to Dark Matter interactions~\cite{Ono2022, Reh2021,Cou2020}.  In this light it is important that the ICL frequencies are measured carefully.  
 The \clockT\ and \coolingT\ lines of ytterbium are in common use in many  experiments~\cite{Ono2022, Ger2010a,Gut2018a,Bru2016,Nak2016}.  
 Locating  the  doubly-forbidden  ($\Delta S \neq0$ and $J=0\not\to J=0$ ) clock transition can be challenging without access to a frequency comb and reference.   However, knowledge of the ICL frequency can be used to calibrate a wavemeter and thereby bypass the need for  such items. 

  We are further motivated to test the validity of previous measurements that employed a hydrogen maser as a frequency reference.    Atkinson \ea~\cite{Atk2019}  made use of  absolute frequencies,  but the only published value was for isotope $^{176}$Yb.    In this previous work we carried out isotope shift spectroscopy  on the  $^1S_0$ $-$ $^3P_1$ ICL across all the naturally occurring isotopes (and hyperfine levels).   Careful analysis of the absolute frequencies was not part of the investigation. 
Here, our focus is on the absolute frequencies of the ICL in  \Yb, where we have a carefully assessed clock transition (\clockT) to act as an additional frequency reference. 
Previous measurements of the intercombination line frequencies in ytterbium include  works by ~\cite{Pan2009,Wij1994,Jin1991, Cla1979}, but these all relied on a Doppler broadened lines.  
 
 We have structured the paper as follows:  In Sect.~\ref{ExperimentalMethod} we describe the experimental method beginning with the cold atom preparation, then detailing aspects of the GNSS frequency reference.   A commercial GPS discipline oscillator was used, but since its 10\,MHz reference signal  lacks short term stability, we use it to steer a higher performing quartz oscillator that provides a reference for the frequency comb.   Sect.~\ref{USC} describes the frequency stabilization of the 578\,nm light needed to probe the clock transition, along with the event sequence that is required to carry out the clock-line spectroscopy.    Sect.~\ref{SecResults}
 details results pertaining to the clock line spectroscopy and the intercombination line frequency measurements, before ending with conclusions.

\section{Experimental Method}  \label{ExperimentalMethod} 

\subsection{Atom preparation for spectroscopy} \label{SubsecAtoms}  %

\begin{figure}[h!]
    \begin{center}
        \includegraphics[width=0.49\textwidth]{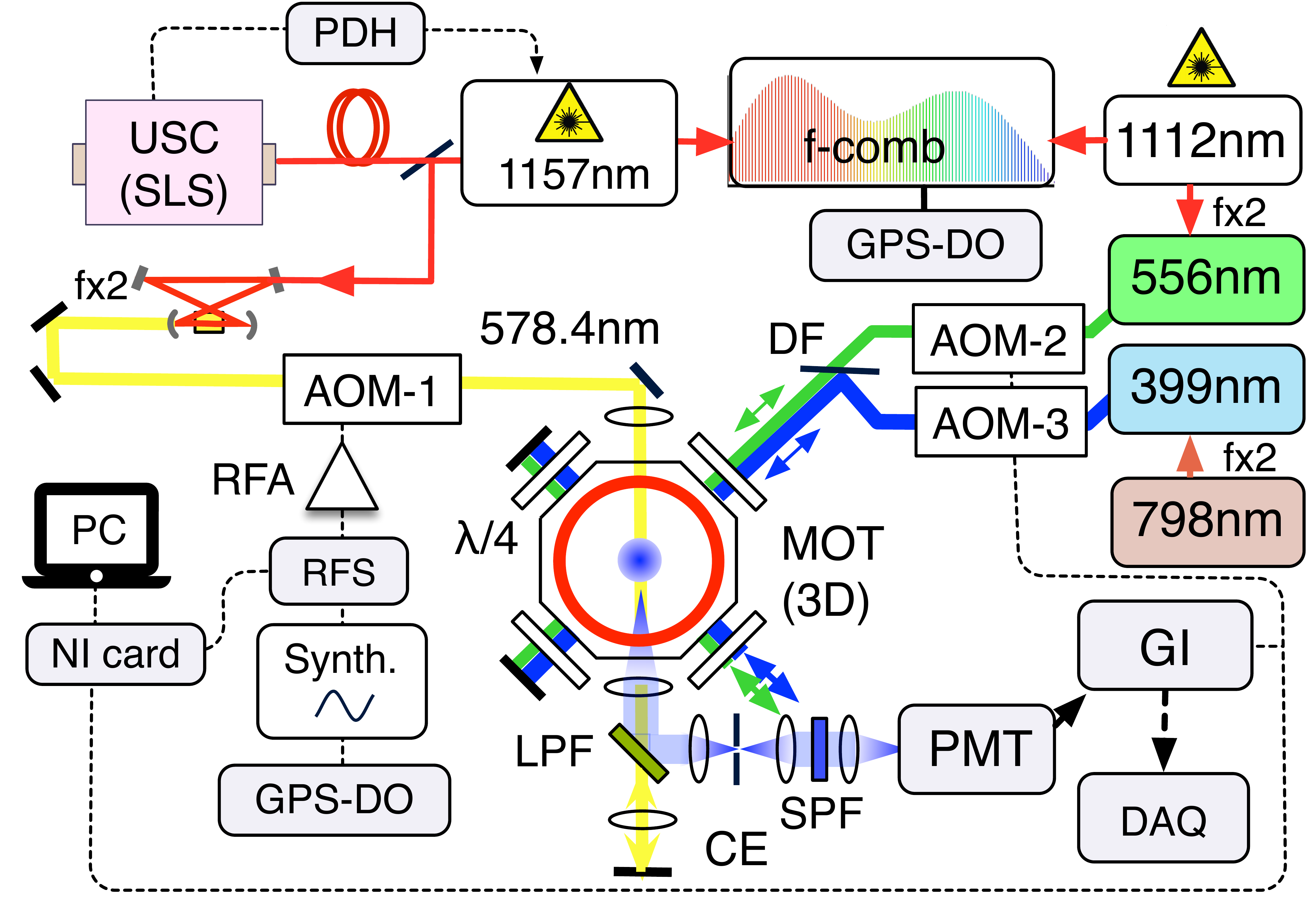} 
        \caption{\small  An outline of the experimental layout for the \clockT\ line spectroscopy in Yb.   The three principal wavelengths  (399\,nm, 556\,nm and 578\,nm) each have their own resonant frequency doubling cavity.  There is a third arm of the dual-wavelength MOT that is perpendicular to  the page.  The 399\,nm and 556\,nm beams co-propagate with the use of dichroic filters and waveplates. A Zeeman slower helps to load atoms into the MOT, which is absent from the figure.  AOM, \AOM;   CE, cat's eye retro-reflector;  DF, dichroic filter; DO, disciplined oscillator; f-comb, frequency comb (in the near-IR);  GI, gated integrator;  LPF, long-wavelength pass filter; MOT, \MOT; NI, National Instruments; PDH, Pound Drever Hall lock; PMT, \PMT;   RFA, RF amplifier;  RFS, RF switch; SPF, short-wavelength pass filter; USC (SLS), ultrastable cavity by Stable Laser Systems.} %
             \label{MainExpLayout}  %
    \end{center}  %
\end{figure}

We begin by describing the clock line spectroscopy that is performed on laser cooled atoms.  The general scheme is illustrated in Fig.~\ref{MainExpLayout}.  
 The MOT operates in two stages: first with 399\,nm light acting on the \fastT\ transition, then 556\,nm light acting on the \coolingT\ transition to reduce the atomic temperature to $\lesssim30$\,\uK.  
  A dual tapered Zeeman slower (crossing zero magnetic field) assists with  loading atoms into the MOT.   The 399\,nm light is generated by frequency doubling 798\,nm light from a Ti:sapphire laser (Coherent MBR-E110) in a resonant cavity.  The 556\,nm light is produced by use of a second frequency doubling cavity, where the incident 1112\,nm light originates from a fiber laser (Koheras)  that injection locks a semiconductor laser (EYP-RWL-1120)~\cite{Kos2015}.  The 399\,nm light is stabilized by  locking to the \fastT\ line in an effused  beam of atoms before they reach a Zeeman slower.
  The 556\,nm light is frequency  stabilized   by 
locking to the \coolingT\ line of the atoms in the same thermal beam using a saturated absorption scheme (but  modified to create an inverted crossover resonance (ICR) that grants higher signal-to-noise of the spectral line~\cite{Sal2017}).   Laser frequencies are made tuneable by the use of  \AOM s (AOMs).

The clock-line laser at 578\,nm is produced by use of a third frequency doubling cavity, where the master laser (at 1157\,nm) is an extended cavity diode laser (LD1001 Time-Base), and   H\"{a}nsch-Couillaud frequency stabilization
is used for the doubling cavity~\cite{Han1980,Nen2016}.  The yellow light frequency is tuned with an additional AOM  (AOM-1 in Fig.~\ref{MainExpLayout}) whose frequency is set with a  synthesizer (Agilent E4428C) that is referenced to a GPS-disciplined oscillator (GPS-DO).   Further discussion of the GPS-DO appears below.  
The optical frequencies of the 1157\,nm and 1112\,nm lasers are found by generating beat signals with a frequency comb.
The optical frequencies of the visible counterparts are evaluated with 
\begin{equation}
\label{EqComb}
\nu = 2 (n f_r + f_o \pm f_b) + f_\mathrm{aom}
\end{equation}
where $f_r$ and  $f_{o}$ are the frequency comb's mode spacing (repetition rate) and offset frequency, respectively ($f_{o}=-20$ MHz), $f_{b}$ is the  beat frequency,  $n$ is the mode number of the comb, and  $f_\mathrm{aom}$ is the radio frequency of the AOM   in the path of the 578\,nm or 556\,nm light (the latter is relevant to the ICL frequency measurements).  The sign of $f_{b}$ may be positive or negative depending on the measurement. 
Some details of the comb have been discussed previously~\cite{Sch2021}, so we make a brief summary here.   Stabilization of the repetition rate occurs at $4f_r=1.00$\,GHz. This avalanche photodetector signal is phase locked to the sum of 20\,MHz from a  direct digital synthesizer  (DDS) and 980\,MHz from a dielectric resonator-oscillator.     The GPS-DO frequency is  doubled to 10\,MHz, which acts as the reference frequency for the DDS, and  to which the \DRO\ is phase-locked. %
The DDS frequency is set so  that the relevant comb beat signals lie at the center of available bandpass filters.  E.g.,
the 1157\,nm beat signal lies at the center of a 30\,MHz bandpass filter, while the 1157\,nm laser is simultaneously locked to an utrastable cavity (described below).   

\subsection{GPS disciplined oscillator} 

\begin{figure}[h!]
    \begin{center}
        \includegraphics[width=0.45\textwidth]{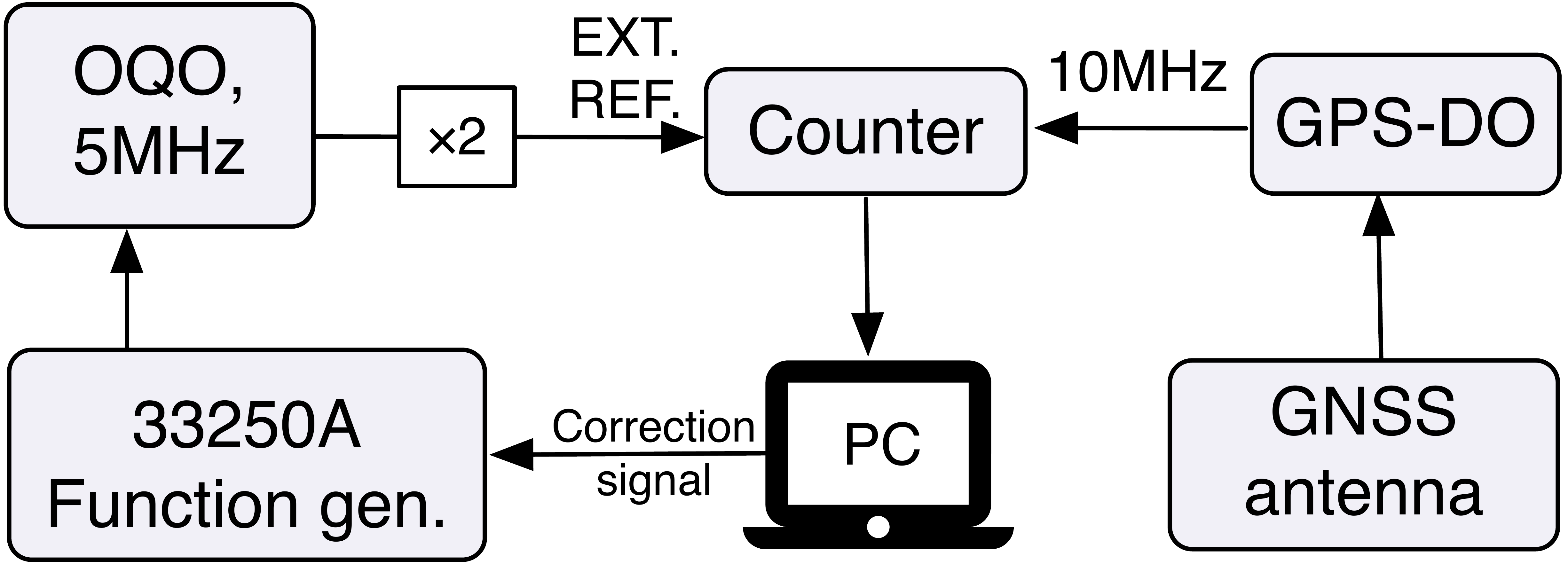} 
        \caption{\small Block diagram of the setup  that locks an Oscilloquartz oscillator (OQO) to GNSS time. The PC records the counter measurements of the GPS-DO's 10MHz signal and performs a simple algorithm for generating a correction signal that is sent to the OQO by way of a function generator in DC mode.  The GNSS antenna is from Antcom Corp. (G5Ant-53AT1).  The OQO is the GPS-DO in Fig.~\ref{MainExpLayout}.  In an alternative scheme we use the PPS signal from the GPS-DO (see text).
        }  %
             \label{GPSServo10MHz}
    \end{center}
\end{figure}

 One may access GPS time with use of a GPS-DO~\cite{Hon2005b}, a GPS-disciplined frequency standard~\cite{Zin2006,Gam2015}, or by a GPS carrier phase link~\cite{Lar1999,Tak2006}, which includes precise point positioning (PPP)~\cite{Pet2015b,Hac2016a,Dub2017}.    
 Our principal frequency reference is that of a GPS disciplined oscillator (GPS-DO), specifically the Trimble Thunderbolt with 10\,MHz and PPS outputs.  However, its 10\,MHz signal only has a stability of $1\times 10^{-11}$ between 1\,s and 100\,s before integrating down.    To provide a better reference for the frequency comb we use 
  an Oscilloquartz oscillator  (OQO-8607-B) that exhibits a fractional frequency instability (FFI) of $2\times10^{-13}$ between 2\, and 500\,s. 
   The frequency of the OQO is tuneable, which allows for  stabilization of the OQO's frequency over longer timescales ($\tau>500$\,s).    This long term stabilization is carried out with the scheme shown in Fig.~\ref{GPSServo10MHz}.
The PC records the counter's frequency measurements of the 10\,MHz signal from the  GPS-DO with 3\,s gate time.  A mean value is determined every 160 samples (480\,s), and the difference from precisely 10\,MHz is used to produce a correction signal that steers the OQO.   The correction signal is provided by a function generator (Agilent 33250A) in DC mode.
The OQO thus becomes the disciplined oscillator.  
The tuning rate of the OQO in terms of the at 578\,nm light frequency (as it is multiplied by the frequency comb)  is 3.1\,kHz/mV 
and the smallest incremental step is 100\,$\mu$V.  
The function generator's DC level ($V_k$) is adjusted in proportion to the offset from 10\,MHz,  denoted $\Delta f$.    The gain (or tuning coefficient) used to correct the OQO is $g=0.5$\,mV/mHz, such that $V_k = V_{k-1}-g \Delta f$ for event $k$.  The time constant of the servo is 4.5\,hrs.  
After an integration time of $5\times10^4$ \,s the square root Allan variance of the counter's measurement is $3.2\times10^{-13}$ and the mean offset from 10\,MHz is 3.0\,mHz  (or 3 parts in $10^{13}$).   
 Hence, the scheme ensures that there is synchronization between the GNSS receiver's internal timing reference and the external clock (i.e., the OQO).
 
 We need to mention that for many of the earlier measurements seen below, a slightly different scheme was employed.  Rather than use the 10\,MHz signal from the GPS-DO, we made use of the PPS (pulse per second) signal.   The counter was set to make 1\,s period measurements instead of counting 10\,MHz.  We have since realized that the counter was producing a gate-time dependent error.
 Only when the gate time of the counter is extended to \si\ 18\,s does the error become insignificant.   The effect created an  offset in the clock line frequency measurements of
 $7\times10^{-11}$ in fractional terms.   This shift has since been accounted for in the inter-combination line data   (further discussion appears below).
 The scheme using the PPS was tested with both a Novatel Propak6 and the Trimble Thunderbolt unit.  They both generated the same $7\times10^{-11}$ offset (for the same gate $+$ dead time of 6\,s). 

  A frequency accuracy at the level of a few parts in $10^{13}$ should be possible with a GPS-DO~\cite{Def2017}.   The difference between the predicted UTC (as GNSS uses) and the post-processed  \textit{true}-UTC published in the BIPM Circular T should not affect the results here, as they are in the $10^{-15}$ range~\cite{Dub2017}

%

\subsection{Laser stabilization and event sequence} \label{USC} 

For the \Yb\ clock line spectroscopy a 1157\,nm laser is stabilized  with use of an ultrastable cavity (USC) and Pound-Drever-Hall lock developed by Stable Laser Systems~\cite{Nen2016}.
   The linear coefficient of thermal expansion is zero at $29.7(2)$\,\degC, and the second order thermal expansion coefficient is $6.6\times10^{-10}$\,K$^{-2}$  (for $\nu=259.1$\,THz).  These were measured with the aid of a hydrogen maser (in our lab). The free spectral range of the cavity is 1496.52\,MHz. 
    By locking to the fundamental   mode of the cavity and offsetting the 578\,nm light  by   $\sim -287$\,MHz with  AOM-1, the \clockT\ transition frequency is reached.   The drift rate  of the 1157\,nm laser locked to the USC is  $+17(1)$\,mHz/s.   This was deduced by comparing the change in the AOM  frequency at the center of the \clockT\ line  over 8 months  (and accounting for the factor of 2).   
   This is also observed by a more direct comparison between the USC stabilised laser and the GNSS signal via the frequency comb, but with  some variation either side of 17\,mHz/s from month to month.   

\begin{figure}[h!]
    \begin{center}
        \includegraphics[width=0.48\textwidth]{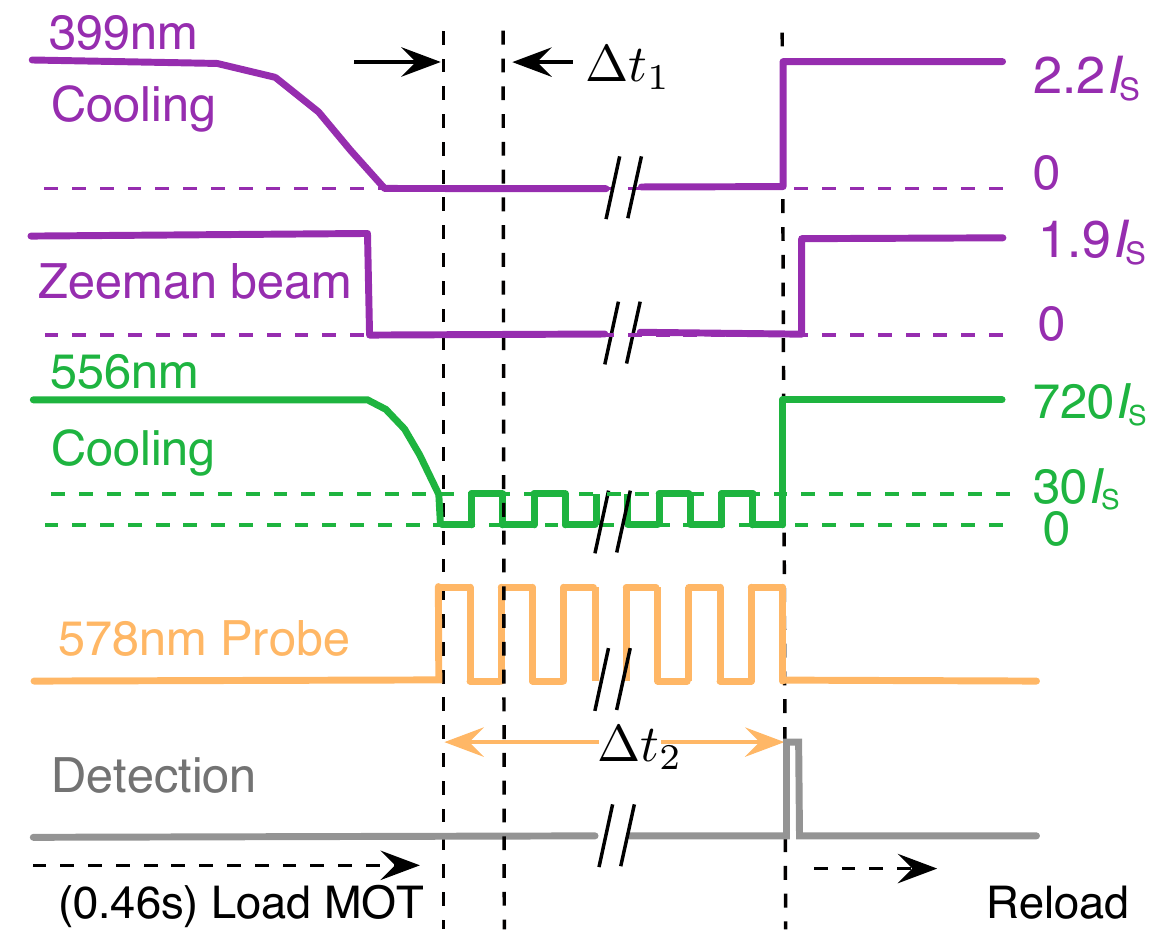} 
        \caption{\small Event sequence of  the light fields for the \clockT\ line spectroscopy.  
       The saturation intensities are $I_S^{(399)}=595$\,W\,m$^{-2}$ and $I_S^{(556)}=1.4$\,W\,m$^{-2}$  for the \fastT\ and \coolingT\ transitions, respectively.  The frequency detuning of the cooling lasers is also varied to optimize signal level.  The Zeeman beam refers to the light used in the Zeeman slower.  Durations  $\Delta t_1 = 1$\,ms and $\Delta t_2 = 40$\,ms.   
        }  %
             \label{SequenceYb}
    \end{center}
\end{figure}

%
The event sequence for the clock line spectroscopy is presented in Fig.~\ref{SequenceYb}.  Each cycle has a duration of 0.5\,s, the majority of which is used to load atoms into the MOT.   The 399\,nm light is ramped down over 20\,ms  during which time the second stage cooling with 556\,nm light takes effect. The 556\,nm intensity is also ramped down over a similar timescale to reduce the temperature of the atoms to $<30$\,\uK~\cite{McF2018}. 
  The magnetic quadrupole field  is held fixed with a $z$-axis gradient of 0.42\,T$\cdot$m$^{-1}$ (the $z$-axis is into the page of Fig.~\ref{MainExpLayout}).  To avoid light-shifting the atoms the 556\,nm and 578\,nm light pulses are interleaved with a duration of  600\,\us\ for  the yellow light and 400\,\us\ for the green light (for later measurements it is 500\,\us\ for both: the difference is not significant).  This interleaving lasts for a period of 40\,ms,  
 after which the  399\,nm is switched back on and detection of the ground state atoms is made (and MOT loading recommences).  
 The maximum intensity of the 578\,nm light was 39\,kW$\cdot$m$^{-2}$ with a corresponding Rabi frequency, $\Omega_R$, of \si11\,kHz (representing \si\ 1/20$^{th}$ of the  transition linewidth).  
Fluorescence at 399\,nm is filtered both spatially and spectrally before detection with a photomultiplier tube (PMT, Hamamatsu H10492-001)~\cite{Kos2014}.  The PMT signal is received by a gated integrator (SRS SR250), which outputs to a \DVM\
for data acquisition.  
The gated integrator is triggered by the main control sequence program so that detection is made within 300\,\us\ of the 399\,nm  light  resuming (the SNR of the clock line spectrum is not  sensitive to this value).  The gate width was set to a maximum of 15\,\us. 
The interleaved pulse technique has previously been applied to \clockT\ lines in strontium~\cite{Cou2003a}, ytterbium~\cite{Hoy2005}, and mercury~\cite{Pet2008}. 

\section{Absolute frequency measurements}  \label{SecResults}

\subsection{Clock-line spectroscopy} \label{ClockLineSpec} 

\begin{figure}[h!]
    \begin{center}
        \includegraphics[width=0.49\textwidth]{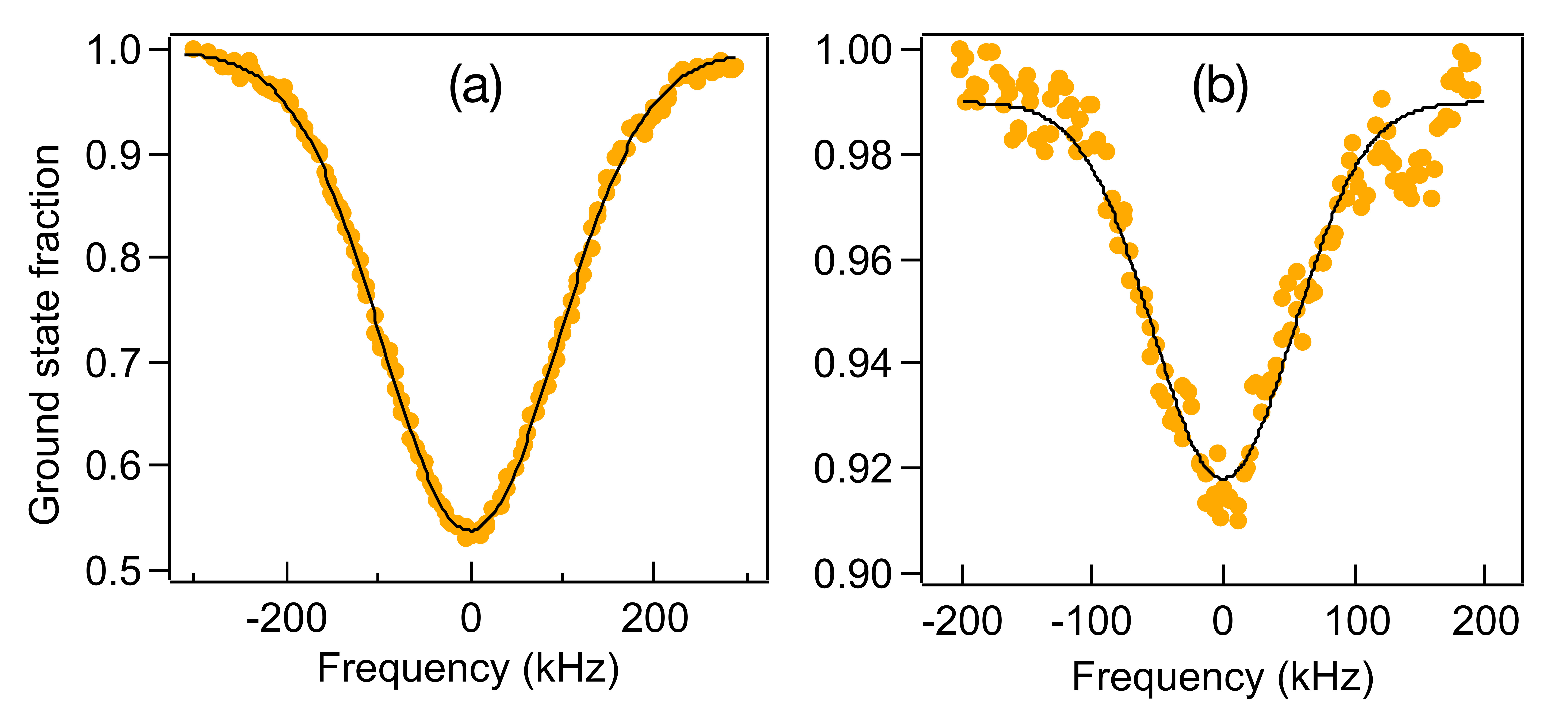}  
        \caption{\small   Ground state fraction versus frequency for the \Yb\ \clockT\ transition taken with a Rabi frequency of  (a) 11\,kHz and  (b)  0.9\,kHz. The curve fits are gaussian line shapes.  The full-width at half maxima are 220\,kHz and 130\,kHz, respectively.  }  %
             \label{FigDoppler}  
    \end{center}
\end{figure}

The majority of the clock-line frequency measurements reported here rely on line fitting to the Doppler profile produced by the cold cloud of slow moving Yb atoms (when shelved  to the $^3P_0$ level by the 578\,nm probe light).  An example of the spectrum is shown in Fig.~\ref{FigDoppler}(a).   This was recorded with a Rabi frequency of \si 11kHz.  About 45\,\% of the atomic population in the ground state is depleted.  From the Gaussian line shape fit the FWHM is 220\,kHz. The time taken to generate the trace was 80\,s.    A Gaussian line shape was used expecting the  spectrum to be thermally (Doppler) broadened.  However, this is not completely the case.  Figure~\ref{FigDoppler}(b) shows a spectrum taken with a much lower intensity (Rabi frequency = 0.9\,kHz).  The FWHM here is 130\,kHz.  Assuming Doppler broadened, this corresponds to a temperature of 20\,\uK, which is consistent with temperature measurements relying on ballistic expansion~\cite{Sal2017}.  For all the clock line frequency measurements reported here we used the higher intensity with higher SNR.  There was no observable light shift at the different intensities, as discussed below.  Our Rabi frequency estimates  rely on a natural line width estimate of 43\,mHz from Derevianko and Porsev~\cite{Der2011a}, used in the expression $\Omega_r =\gamma \sqrt{I/2I_s}$, where $\gamma$ is the natural inewidth, $I$ is the intensity and $I_s=2.9\times10^{-7}$\,W\,m$^{-2}$ is the saturation intensity. 

\begin{figure}[h!]
    \begin{center}
        \includegraphics[width=0.48\textwidth]{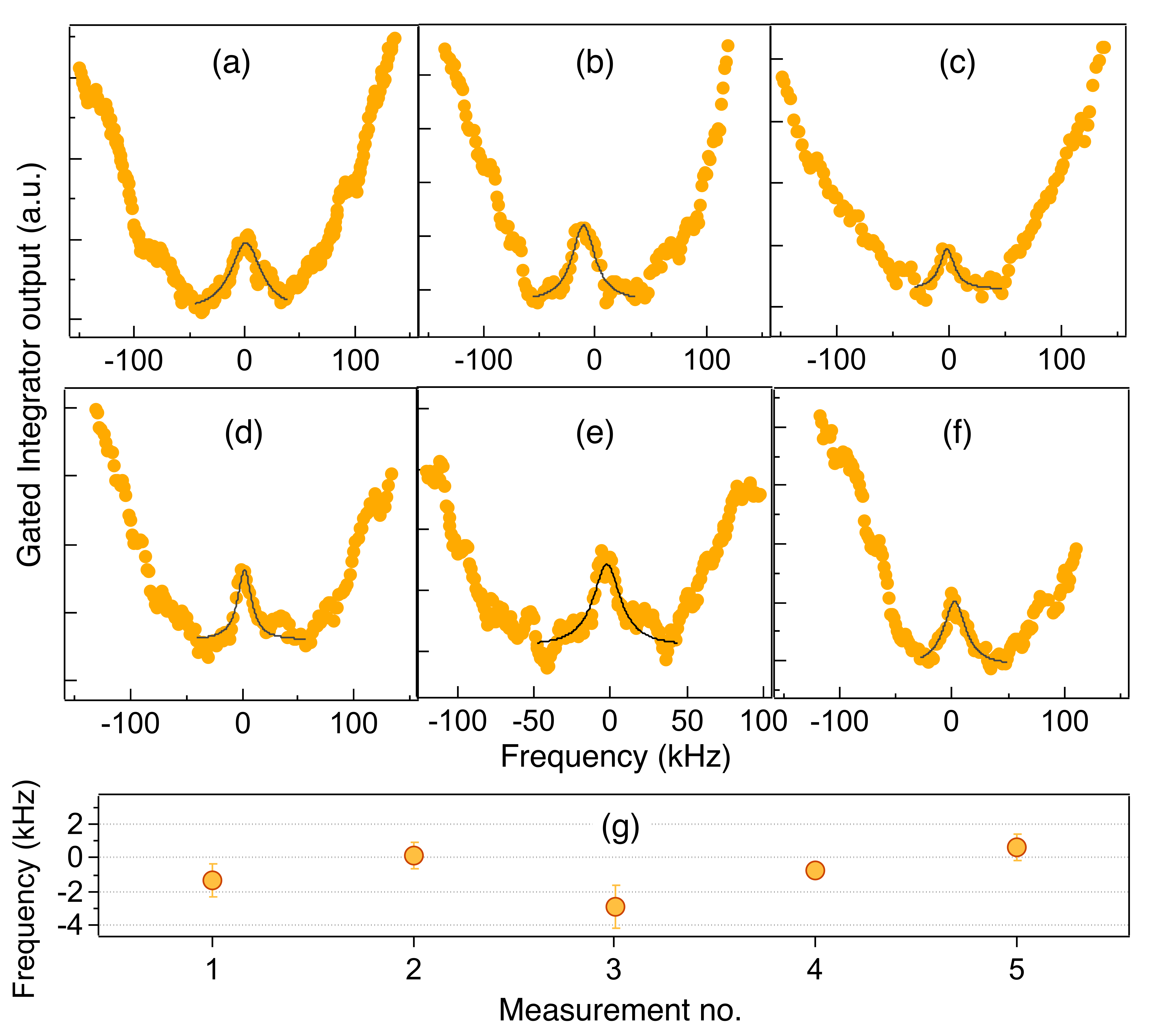} 
        \caption{\small (a) to (f) Examples of  spectra of the \clockT\ (clock) line showing the saturation feature. 
        The vertical axis is proportional to the number of atoms in the $^1S_0$ ground state. 
          (g)  Frequency of the line center for repeated measurements.  The frequency is offset by the AOM-1's frequency  (287\,MHz).  The error bars are a statistical uncertainties from the lineshape fits.  
        }  
             \label{Clockline}
    \end{center}
\end{figure}

For all measurements of the \clockTdash\ transition frequency  the 578\,nm light   was retro-reflected with a cat's eye (lens-mirror combination).  About 60\%  of the 578\,nm light reaches the atoms on the return pass (the loss being due to a long-pass filter used to reflect 399\,nm fluorescence to the PMT).   Under optimal conditions one may expect to see a recoil doublet, due to the atom's recoil when absorbing/emitting a photon~\cite{Pet2008}.  For  the clock transition in \Yb\ the recoil peak separation should be 6.98\,kHz, which is just below the current resolution.   Instead  we observe a single saturated absorption dip (such a dip  with Yb was mentioned by Hoyt \ea~\cite{Hoy2005}).  Examples are shown in Fig.~\ref{Clockline}.   The time taken for the sweep varied from 80\,s for Figs.~\ref{Clockline}(c) to \ref{Clockline}(f),  100\,s for Fig.~\ref{Clockline}(b),  to 160\,s for Fig.~\ref{Clockline}(a).    %
The curve fits are Lorentzian line shape functions. The mean of FWHM of the saturation features shown here is  \si 23\,kHz. 
 There is some line width dependence on 578\,nm light intensity, but for frequency measurements we use the maximum light available to enhance the SNR.
  The background Doppler profile at the time of these measurements was \si300\,kHz full-width at half maximum, only the base of which is shown here. 
  An important aspect is that the saturation dip lies near the center of the Doppler profiles.  These saturation dips were only observed when the yellow and green light was interleaved during the probe period.
A set of frequency measurements based on  saturation dips (or peaks in this case) is shown in Fig.~\ref{Clockline}(g).
 There was approximately 4 minutes  between measurements.  The statistical (1-$\sigma$) variation here is 1.4\,kHz. 
 Unfortunately, after a time this saturation feature could not be reproduced on demand.  The main experimental change since these traces were recorded is that  the number of atoms has more than halved due to a significant reduction in 798\,nm laser power.   Despite this, the SNR of the Doppler background (e.g.  Fig.~\ref{FigDoppler}a) has improved to the point where the statistical uncertainty of  the line-center frequencies is comparable with that seen here.   The number of atoms contributing to the clock line signal is now \si\, $2\times10^5$.  
 
\begin{figure}[h!]
    \begin{center}
        \includegraphics[width=0.49\textwidth]{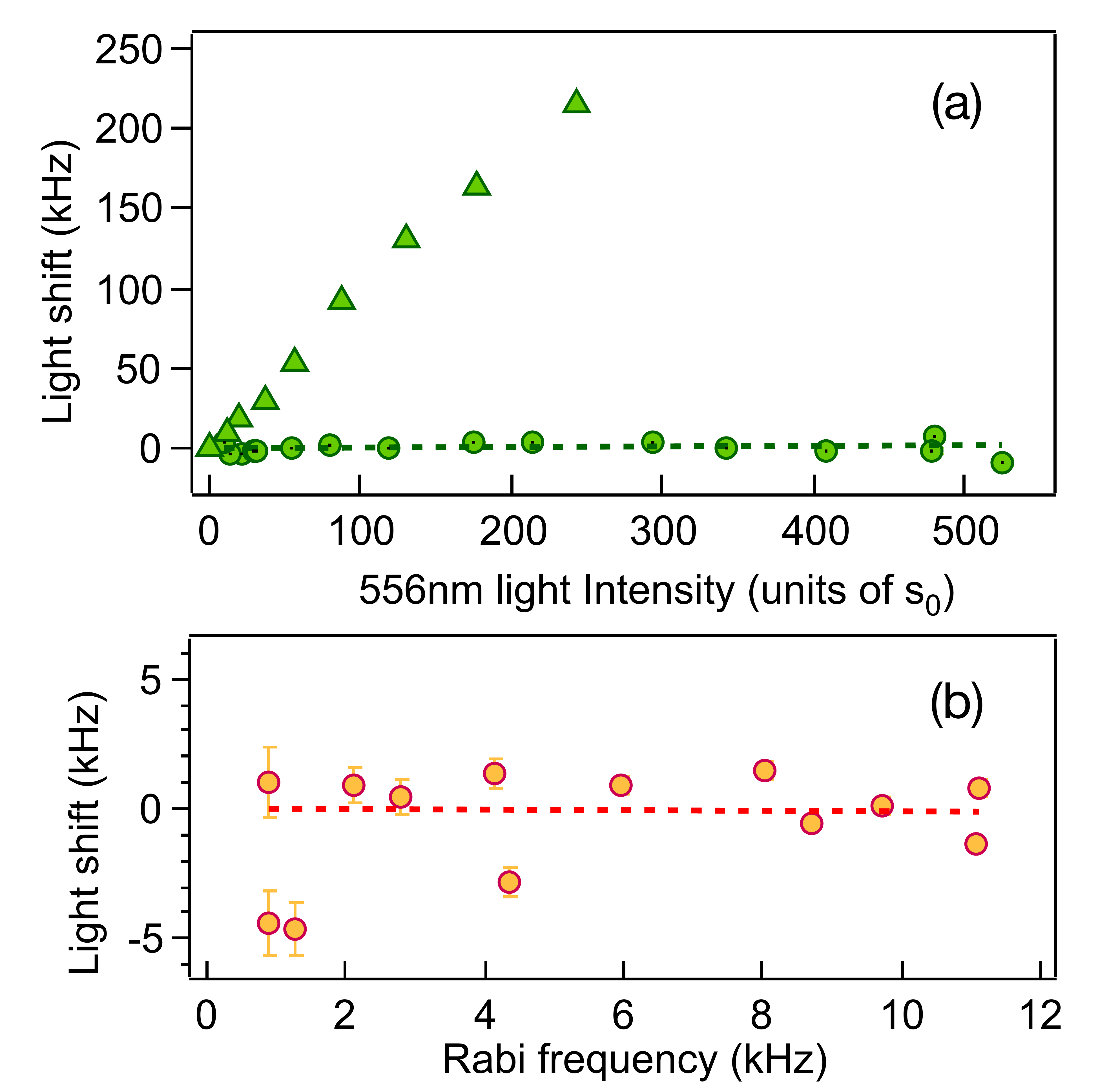}  
        \caption{\small  (a)  AC Stark shift of the clock transition (\clockT) versus the intensity of the 556\,nm light used in the second stage MOT. ($\triangle$)  Single 40\,ms pulses of 556\,nm light and 578\,nm are applied simultaneously.  ($\odot$) The 578\,nm light and 556\,nm light are interleaved over a period of 40\,ms with 500\,\us\ pulses.    $s_0 = I/ I_S^{(556)}$. 
        (b) AC Stark shift of the clock transition versus the intensity of the 578\,nm light (in terms of  Rabi frequency).
 }  
             \label{LightShift}
    \end{center}
\end{figure}

    The \clockT\ transition (with a lifetime of \si\,4\,s) has very low sensitivity to  environmental factors~\cite{Por2004a} and most systematic shifts are negligible at kilohertz resolution; however, care must be taken to ensure the atoms are not exposed to laser light during the excitation phase(s). 
     The 399\,nm light is blocked with a physical shutter (Cambridge Technology Inc.) during the 578\,nm probe.   For the 500\,\us\ probe pulses the 556\,nm light is not physically blocked but switched off via  AOM-2.  The residual light intensity is $0.04 I_S^{(556)}$. 
     We find that this produces negligible shift at the  kilohertz level.  
   In support of this we present light shift measurements due to the 556\,nm MOT beams in Fig.~\ref{LightShift}(a) $-$ looking initially at the data points marked with triangles.  
   Rather than interleave the 556\,nm and 578\,nm light, a full 40\,ms pulse of 578\,nm is applied and different levels of 556\,nm light are imposed on the atoms.  Only at zero intensity are the two beams interleaved.   The light shift dependence is approximately linear. 
Of importance is that  there is no discontinuity between the zero-intensity case and the extrapolation of the  bulk data. Thus, we can be assured that the residual 556\,nm light is producing negligible shift. 
The data points with  marked with circles in Fig.~\ref{LightShift}(a) show the result of changing the green light intensity when using the interleaved method.  The line of best fit has a slope that is almost zero.   For the clock line spectroscopy we operate at a 556\,nm beam intensity of $s_0=30$, which has a corresponding shift of +0.2\, kHz.  
For completeness we also present the AC Stark shift of the \clockT\ transition as a function of the 578\,nm intensity, where the intensity is expressed in units of Rabi frequency (Fig.~\ref{LightShift}b).    The slope is zero within the measurement uncertainty. 
At the lowest intensity the FWHM reduced to 130\,kHz (as seen above).   

The gated integrator averages over 3 samples (with an exponential moving average). The leads to  a hysteresis-like effect on the line profiles causing a line-center shift.  To counter this we scan in both directions and take the mean of the two for the line-center frequency.   For example,  for or a scan across 400\,kHz in 80\,s the corresponding shift is 8.5(0.3)\,kHz.  

\begin{figure}[h!]
    \begin{center}
        \includegraphics[width=0.5\textwidth]{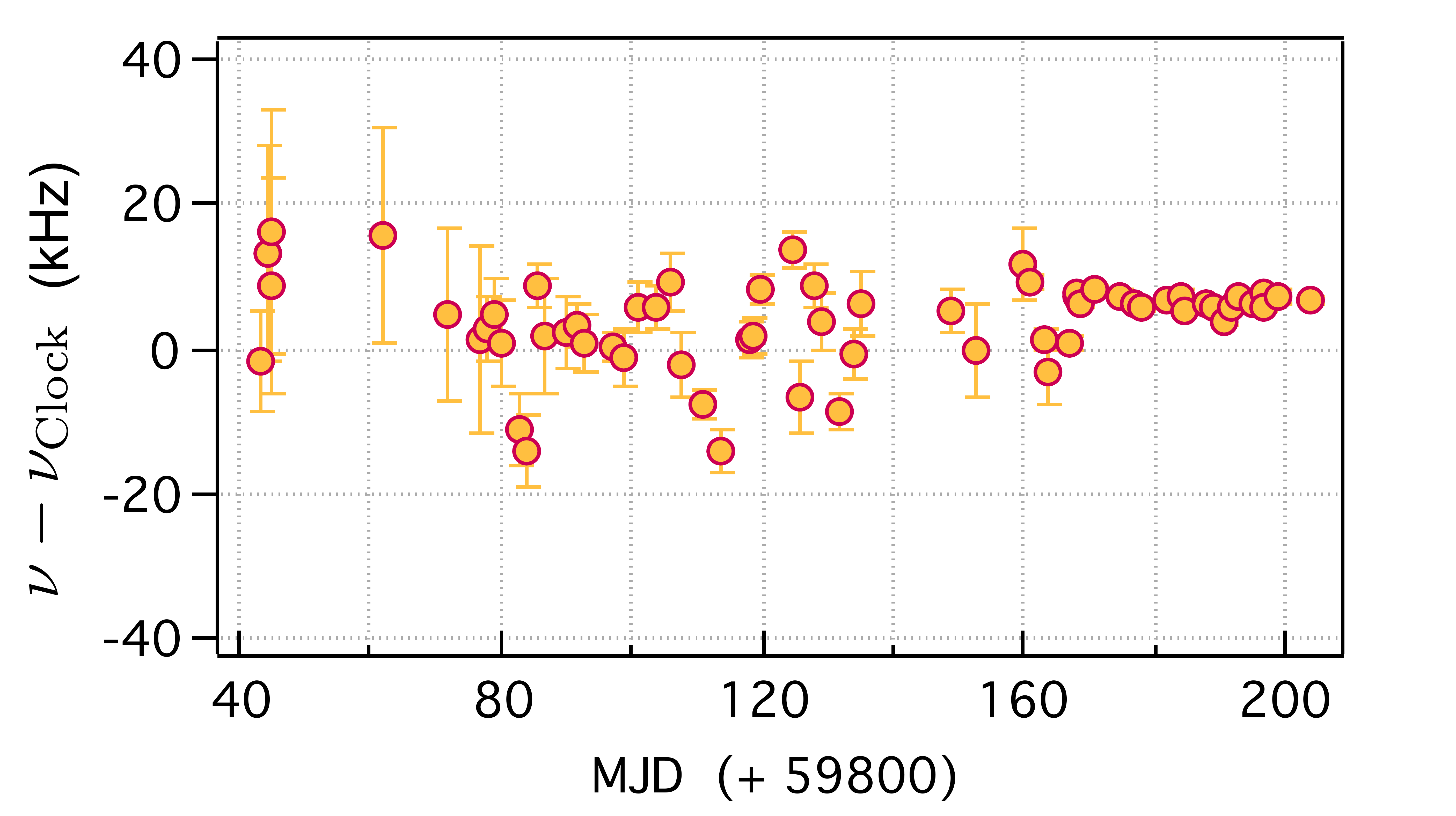} 
        \caption{\small  Absolute frequency measurements of the \Yb\ $^1S_0$~$-$~$^3P_0$\, line.  The frequency is offset  with respect to the absolute frequency of the transition $\nu_\mathrm{Clock} =  518\,295\,836\,590.8636$\,kHz~\cite{Ybclockline2021}.           
   }  
             \label{Clockfreq}
    \end{center}
\end{figure}

Our absolute frequency measurements of the \clockT\  transition in \Yb\ are shown in Fig.~\ref{Clockfreq}(a).   The frequencies are offset by the recommended BIPM value of  $\nu_\mathrm{Clock} =  518\,295\,836\,590.8636$\,kHz~\cite{Ybclockline2021}, which follows from the work of several research institutes~\cite{Nemitz2016,Piz2017,McG2019}.
 The weighted mean offset  from the  well established frequency is 5.9(0.9)\,kHz (or $1.1\times10^{-11}$ in fractional units).   
An offset is to be expected when relying on the Doppler line shapes.   
 During the excitation pulses the atoms are in free fall.  The change in their velocity due to gravity over 500\us\ has a corresponding mean Doppler shift of  4.2\,kHz~\cite{Hoy2005}.  Since the downward beam has more intensity than the reverse beam, it plays a greater role in forming the line shape.  This downward beam frequency is redshifted with respect to the falling atoms (hence the measured frequency is slightly higher than expected).  
 The 4.2\,kHz shift does not account fully for the 5.9\,kHz offset, but it lies within 2-$\sigma$. The difference corresponds to 3.2 parts in $10^{12}$. 

Other perturbations to the optical transition  frequency were explored.
An examination  was made of  AC Stark shifts; for example, from  the 556\,nm beams as described above.  %
There was some evidence that the far-detuned (-7\,$\Gamma $) Zeeman slower beam was producing a light shift.  To mitigate  this the Zeeman slower beam was blocked (with a physical shutter) during the 578\,nm probe.   This was implemented at day 164.  One sees that the frequency variations have reduced beyond this time.  
As well as inserting the shutter, the Zeeman beam was weakly focussed giving rise to a larger number of atoms in the MOT, thus increasing the SNR and reducing the line-center uncertainty.   In retrospect, the Zeeman beam appears to have produced very little light shift (comparing data before and after day 164). 
 Shifts with respect to a bias magnetic field were tested, but as expected for the $J=0$ transition there was no observable shift.   
 
Systematic shifts were investigated on the side of the GNSS frequency reference.  These included:
  (i ) the  cable length between GNSS antenna and receiver, (ii) the cable length between   receiver and counter,  and (iii) receiver temperature.  No shifts were found at the kilohertz level. 
   A check was also made with regard to the frequency comb.  The integrity of the DDS was tested by changing its reference frequency from 10\,MHz to 220\,MHz.  The 220\,MHz was provided by a synthesiser referenced to the OQO.  Shifts in the 1157\,nm comb beat frequency were less than 100\,Hz (in the noise). 


\subsection{Intercombination line measurements} \label{ICLSpec} 

The link with GNSS time  provides a reliable means of measuring the intercombination line frequencies in Yb (in conjunction with the Yb clock line).  
Here we focus on two lines:  $^{1}S_{0}-\,^{3}P_{1}$ ($F=1/2$ and $F=3/2$).   We will denote the two lines as $ICL_{1/2}$ and $ICL_{3/2}$, respectively. The natural linewidth of this transition is 184\,kHz~\cite{Bel2012}.   We point out that there is a slight difference in the experimental approach between the two lines.  
Both are methods with sub-Doppler resolution.
For $ICL_{3/2}$ we use the more typical saturated absorption scheme where the background magnetic field is cancelled so all the upper $m_F$ levels are degenerate.
   For $ICL_{1/2}$  we use a small bias field  (0.09\,mT) to generate an inverted crossover resonance (ICR).  In both cases the light polarization is parallel to the magnetic field so that $\Delta m_F=0$, and third harmonic detection is used in the  signal demodulation. 
   The saturated absorption spectroscopy (SAS) and ICR schemes make use of a thermal Yb beam leading to the Zeeman slower and MOT.  Laser light at 1112\,nm is frequency doubled in a resonant frequency doubling cavity producing 556\,nm light for the saturation spectroscopy (as well as for second stage atomic cooling in the main clock chamber).  Further details of the  SAS and ICR setups may be found in \cite{McF2016, Atk2019}.   The fundamental light (1\,mW) is combined with light from the frequency comb on an avalanche photodiode, generating a beat note with frequency $f_b$.  The comb light is bandpass filtered with a  15\,nm FWHM filter.  
    Based on the parameters of the frequency comb ($f_o= -20$\,MHz, $f_r$ and $n$ depend on the line), and the  frequency of  AOM-2 (108.80\,MHz) that tunes the 556\,nm light frequency, we can determine the absolute optical frequency of the transition by use of Eq.~\ref{EqComb}. 

   The dominant systematic shifts and their uncertainties are listed in Table~\ref{Tab:Uncert} for the two lines.  The central column is data for  $ICL_{1/2}$  (the \ICR).  It experiences smaller shifts with regard to beam alignment and the cat's eye lens position compared to  $ICL_{3/2}$; furthermore,   $ICL_{1/2}$  has no discernible shift with respect to the magnetic field at the kilohertz level due to the nature of the resonance~\cite{Sal2017}.   The only systematic that requires a correction is the AC Stark shift (values are stated in the Table~\ref{Tab:Uncert} heading).  
    The cat's eye provides parallelism between the forward and reverse beams and reduces the sensitivity to beam alignment.   Shifts associated with horizontal beam alignment are negligible, but there remains sensitivity to vertical beam alignment.

    \begin{table}  
    \caption{\label{Tab:Uncert}Uncertainties for the \Yb\  \coolingT\ $F=1/2\rightarrow 1/2$ and  $F=1/2\rightarrow 3/2$ transitions. There are corrections of +1.8\,kHz and +10.4\,kHz due to the light shift for the two lines, respectively. }
    \begin{tabular}{lcc}
    \hline
        Effect &  $\Delta u(ICL_{1/2})$ & $\Delta u(ICL_{3/2})$ \\ 
         &  (kHz) &  (kHz) \\
        \hline
        Cat's eye lens position  & 11  & 47  \\
        Cat's eye mirror position  & 8 & 8   \\
     AC Stark shift   & 2 &  3 \\
      1st order Zeeman  & 0 &  4  \\
      Servo error & 3 & 3   \\
      Vertical beam alignment  & 31 & 48 \\
        \hline
        Total   & 34  &68   \\
        \hline
\end{tabular}
\end{table}

\begin{figure}[h]  
    \begin{center}
        \includegraphics[width=0.5\textwidth]{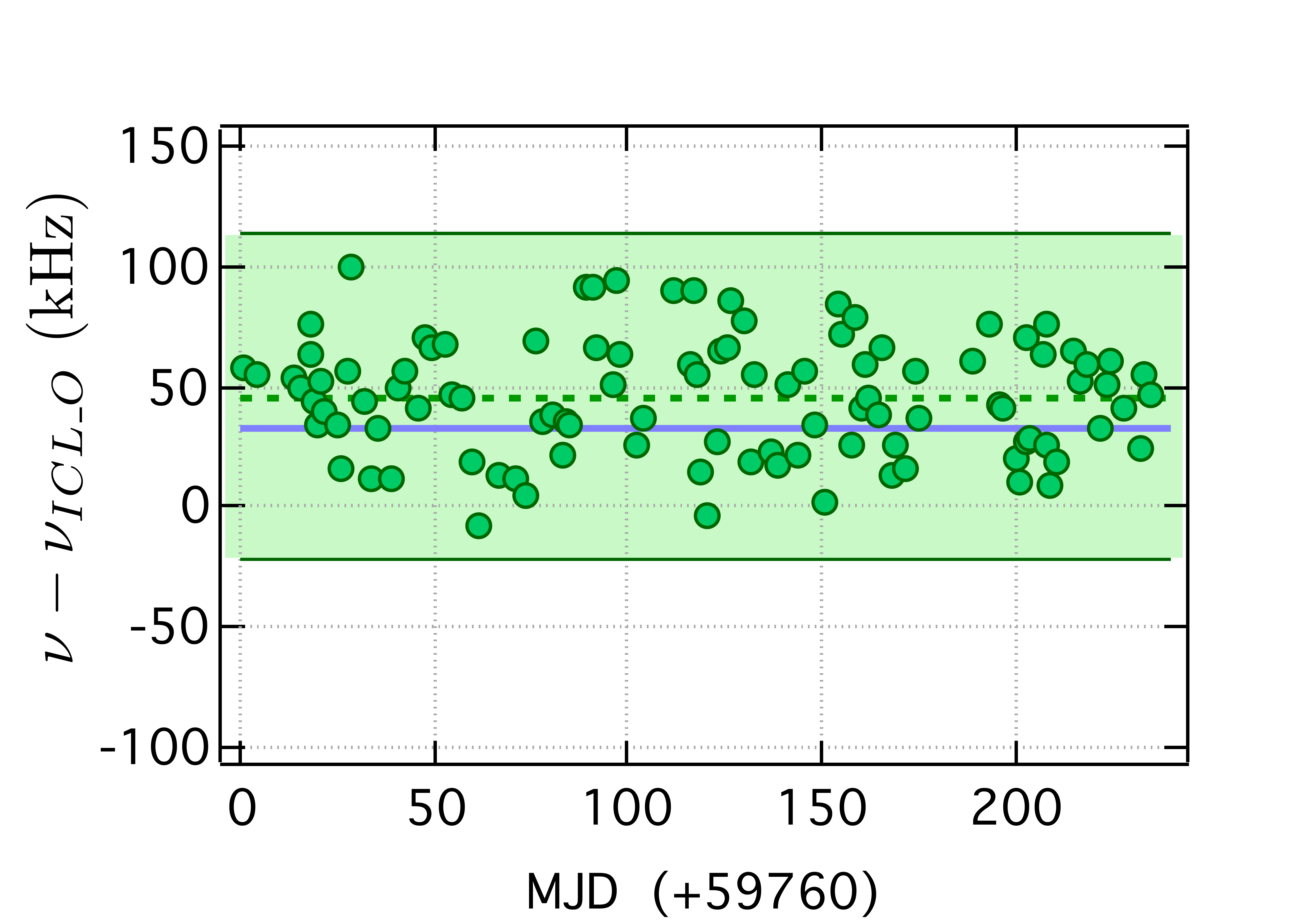} 
        \caption{\small Absolute frequency measurements of the \Yb\ $^1S_0$~$-$~$^3P_1$\,($F$=3/2)  line.  The frequency is offset by $\nu_{ICL\_O} = 539\,390\,405\,700$\,kHz.  The solid line shows a 2019 measurement taken using a H-maser as the frequency reference. 
         The dashed line represents the current  mean frequency taken \wrt\ GNSS time and the  shaded region represents the  1-$\sigma$ systematic  uncertainty. 
        }  
             \label{MeasuresSAS}
    \end{center}
\end{figure}

Absolute frequency measurements of the \Yb\  $^1S_0$~$-$~$^3P_1$ $(F=3/2)$  transition, recorded  over  \si240 days, are  shown in Fig.~\ref{MeasuresSAS}.  For the majority  of this data a correction has been applied due to the counter inaccuracy mentioned above.  At the clock line frequency the shift between the two GPS servo methods was $-31.7$\,(2.0)\,kHz.  The ICL frequencies have therefore been reduced by 33.0\,kHz (it scales by the ratio of the optical frequencies). 
  The thick solid line shows a measured value from 2019, taken using a hydrogen maser as the main frequency reference (the absolute values were used in Ref.~\cite{Atk2019}, but they were not published).   The measurements taken with  the GNSS reference are  in good agreement with the previous measurement. 
 Both the data in Fig.~\ref{MeasuresSAS} and the 2019 measurement include a $-10.4$\,kHz light shift.  Accounting for this, 
 the  mean frequency is   539\,390\,405\,756.0\,(2.4)$_\mathrm{Stat}$\,(68.0)$_\mathrm{Sys}$\,kHz.    The systematic  uncertainty is represented by the shaded region in Fig.~\ref{MeasuresSAS}.   The statistical (systematic) uncertainty corresponds to 1.3\,\% (37\,\%) of the natural line width of the transition.  

\begin{figure}[h!]
    \begin{center}
        \includegraphics[width=0.5\textwidth]{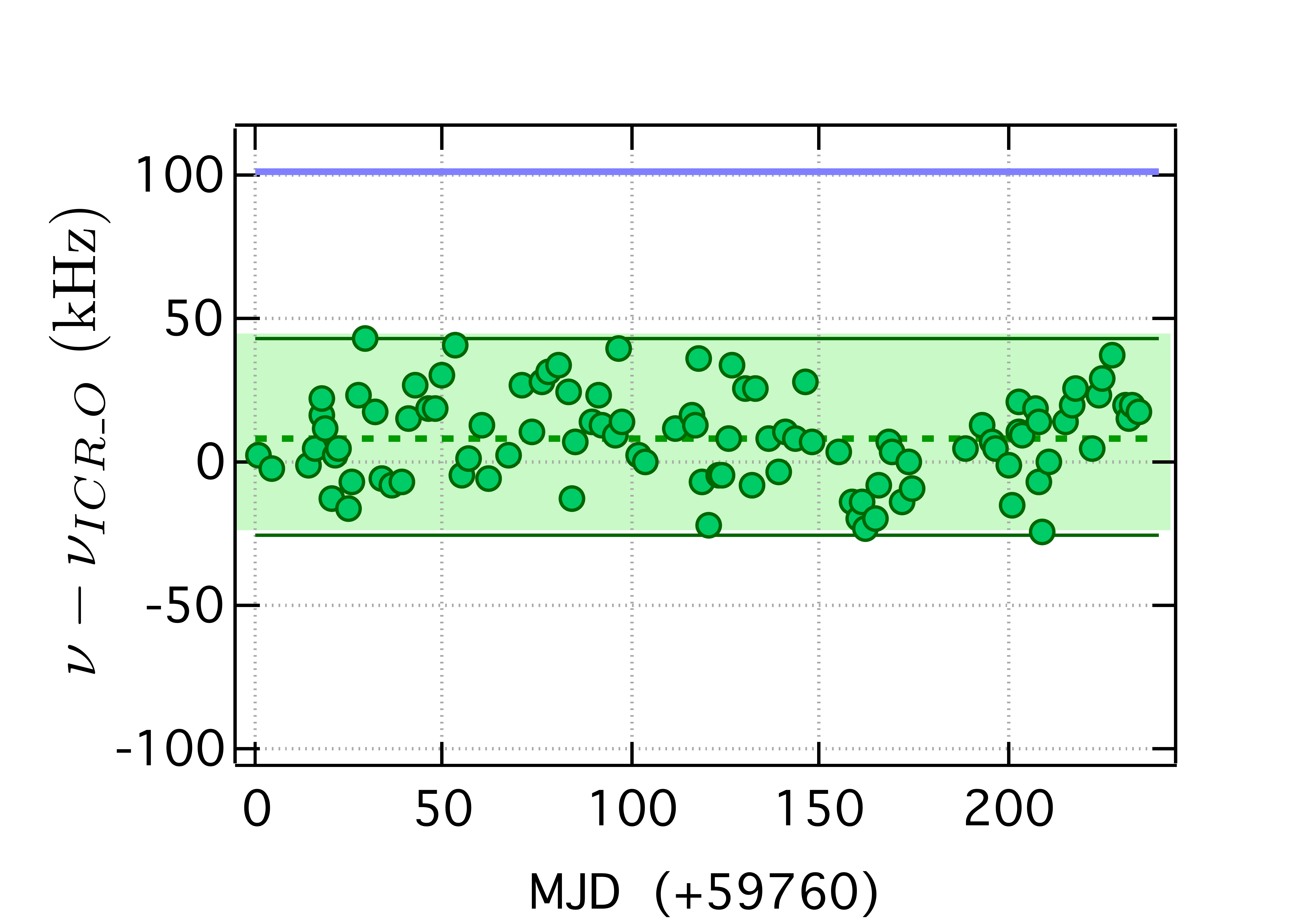} 
        \caption{\small Absolute frequency measurements of the \Yb\ $^1S_0$~$-$~$^3P_1$\,($F$=1/2) line.  The frequency is offset by $\nu_{ICR\_O} = 539\,384\,469\,000$\,kHz.  The solid line shows a  measurement  taken in 2019 using a  H-maser as the  frequency reference.  
        }  
             \label{MeasuresICR}
    \end{center}
\end{figure}

Similar measurements were performed for the \Yb\ $^1S_0$~$-$~$^3P_1$\,($F$=1/2) line as summarised in Fig.~\ref{MeasuresICR}.
     The  frequency is offset by $\nu_{ICR\_O} = 539\,384\,469\,000$\,kHz,  where ICR implies it is an inverted crossover resonance.
 The statistical variation is slightly less than that of   $ICL_{3/2}$, 
 which  relates to the observation that  $ICL_{1/2}$  
  is less susceptible to beam pointing variations than $ICL_{3/2}$ 
  (the former relies on a larger subsection of atoms in the atomic beam $-$  or rather, two ever-so-slightly diverging sets of the atoms~\cite{Sal2017}).
There is an offset from a 2019  measurement taken with the H-maser.  It is attributed to a ground-loop that created a  voltage offset at the input of the integrating operational amplifier (op-amp) used to lock the 1112\,nm laser.  A sign switch preceded the integrator and the effect was only manifest for the non-invert condition, hence, it only affected the ICR and not the SAS, which have opposite slope discriminators.  
The input offset of the integrating op-amp causes a shift of -1.51(8)\,kHz/mV in the infrared.   
It can be nulled to within a 1\,mV producing a systematic uncertainty in the visible of 3\,kHz.
 Our estimate for the \Yb\ $^1S_0$ $-$ $^3P_1$ $(F=1/2)$ line  frequency is  539\,384\,469\,010.1~(1.6)$_\mathrm{Stat}$\,(34.0)$_\mathrm{Sys}$\,kHz.     Here the light shift correction is only +1.8\,kHz.  
  %

The reliance on GNSS can be minimised by  considering the ratios of the optical frequencies when systematic shifts of the clock  line frequency are  mitigated~\cite{Ohm2020}. 
 When the beat frequencies for any two lines  are measured with the comb set to the same repetition  frequency, then the frequency ratio  between the lines can be written as

\begin{equation}
\label{EqRatio}
\frac{\nu_1}{\nu_2} = \frac{n_1}{n_2}+\frac{\epsilon_1}{n_2 f_r} - \frac{n_1\epsilon_2}{n_2^2 f_r}+\mathcal{O}^{(2)}
\end{equation}
where

\begin{equation} 
\label{ }
\epsilon = f_o\pm f_b+ f_\mathrm{aom}/2
\end{equation}
The values of $f_b$ and $f_\mathrm{aom}$ are different for the two lines, hence the two values of $\epsilon_1$ and $\epsilon_2$ (the AOM frequencies are divided by 2 when computing $\nu_{1,2}$ in the infrared).  The relative magnitude of the $2^{\mathrm{nd}}$ and $3^{\mathrm{rd}}$ terms of Eq.~\ref{EqRatio} are  parts in $10^7$, so the influence of the GPS-DO through $f_r$ becomes insignificant.   The mode numbers for the green ($F=1/2 \rightarrow 3/2$) and yellow lines are $n_1=1078780$ and $n_2=1036592$, respectively, giving $n_1/n_2= 1.04069875$. With regard to the frequency comb, the main change   made is that the DDS is held at 20\,000\,550\,Hz for all the beat frequency measurements.  This moves the 1157\,nm beat to -60.5\,MHz (from 30.2\,MHz). 
 Measurement of the first order corrections over a series of 40 daily measurements  gives  $\nu_1/\nu_2= 1.040699862276(9)$. This is a separate set of measurements from those of Fig.~\ref{MeasuresSAS} and   Fig.~\ref{MeasuresICR}.  
 With use of the BIPM value for  $\nu_2$ (the clock line) we find the frequency of the \coolingT\ ($F=1/2 \rightarrow 3/2$)  line to be 539\,390\,405\,768.5~(4.5)$_\mathrm{Stat}$\,(68.0)$_\mathrm{Sys}$\,kHz.
 By a similar process, for the ($F=1/2 \rightarrow 1/2$)  line the mean frequency is 539\,384\,469\,005.2\,(4.2)$_\mathrm{Stat}$\,(34.0)$_\mathrm{Sys}$\,kHz.
 The light shift corrections have been applied post multiplication of the clock line frequency.  
The frequency for $ICL_{3/2}$ is 12.6\,kHz  higher than that obtained using the direct frequency comb measurement.  For $ICL_{1/2}$  the difference is only $4.9$\,kHz.    Both differences are well within the systematic uncertainties.

 Given the  frequencies of the two \Yb\  \coolingT\ hyperfine lines we can determine their hyperfine separation. For this purpose we 
 will take a weighted mean (based on the statistical uncertainties) of the two values  measured by the two methods discussed above.  The resultant frequencies are therefore:  539\,390\,405\,761 (69)\,kHz for $ICL_{3/2}$ and 539\,384\,469\,009 (35)\,kHz for $ICL_{1/2}$.
   The hyperfine separation  is found to be 5\,936\,749\,(41)\,kHz.  The uncertainty here is not a simple RMS sum because most of the systematic shifts are partially common between the two lines. 
The hyperfine constant in \Yb\ is two thirds of the hyperfine separation, giving  $A(^3P_1)= 3\,957\,833\,(28)$\,kHz.
     This is higher than the  value reported by Pandey \ea~by 52\,kHz~\cite{Pan2009}, but well within our overlapping uncertainties.

\section{Conclusions} \label{Conclusion} 

With the knowledge of the absolute frequencies of the   \clockT\ transition frequency and the ICL ($F=3/2$) we  evaluate the difference between them to be  21\,094\,569\,168\,(69)\,kHz.     The uncertainty is dominated by that of the ICL.
  This value may be used for comparison in atomic structure calculations involving Yb \textsc{i} when computing the energies of levels. It can also assist with finding the clock transition should a frequency comb and accurate reference not be available (e.g. with a wave-meter).

To summarize, we have  measured the absolute frequencies of the  \Yb\ $^1S_0$~$-$~$^3P_1$\,($F$=1/2)   and  $^1S_0$~$-$~$^3P_1$\,($F$=3/2) lines, where the GNSS timescale has been used to provide a frequency reference (the final values are stated just above).  The  frequency for the $^1S_0$~$-$~$^3P_1$\,($F$=3/2)   line is consistent with previous (unpublished) measurements that relied on a hydrogen maser as the  reference.  The  frequency for the $^1S_0$~$-$~$^3P_1$\,($F$=1/2) line  departed slightly  from a H-maser based measurement for  reasons discussed in Sect.~\ref{ICLSpec}. 
 These measurements provide a new value for the hyperfine constant  $A( ^{171}\mathrm{Yb},\, ^3P_1)$. 
The GNSS reference was validated by carrying out frequency measurements of the \Yb\ clock transition.  The clock line spectroscopy on  laser cooled  Yb atoms  revealed a saturation dip free from the first-order Doppler shift, but for the majority of the measurements the  \si200\,kHz wide Doppler profile was used.   The line-centre frequency agreed with the CODATA value to almost 3 parts in $10^{12}$.  
  Access to the  absolute ICL frequencies of the other ytterbium isotopes can be made via previously measured isotope shifts~\cite{Atk2019,Pan2009}.

\vspace{0.3cm}

 
\flushleft \textbf{Acknowledgment}

We thank E. Ivanov and M. Goryachev for the use of the Oscilloquartz oscillator and C. Blair for the loan of the Trimble Thunderbolt.  We are grateful to N.~Nemitz for  suggesting the ratio approach for optical frequency measurements.  We thank L.~Nenadovi\'c for proofreading the manuscript. 

\textbf{Funding} \ARC\  LE110100054.

\textbf{Disclosures}  The authors declare no conflicts of interest.

\vspace{-0.1cm}


 %

\end{document}